\documentclass[aps,prl,twocolumn,groupedaddress,showpacs,floatfix]{revtex4}
\usepackage{dcolumn}
\usepackage{amsmath}
\usepackage{amssymb}
\usepackage{graphicx}
\usepackage{bm}
\usepackage[T1]{fontenc}
\usepackage{color}
\usepackage{url}
\usepackage[bf]{subfigure}
\usepackage{rotating}

\begin{document}
%
\title{An entropy-based approach to automatic image segmentation of
satellite images}
\author{A.\ L.\ Barbieri$^1$, G.\ Arruda$^1$, F. A. Rodrigues$^1$, O.
M. Bruno$^1$, and L. da F. Costa$^{1,2}$}
\affiliation{$^1$Instituto de F\'{\i}sica de S\~{a}o Carlos,
Universidade de S\~{a}o Paulo, Av. Trabalhador
S\~{a}o Carlense 400, Caixa Postal 369, CEP 13560-970, S\~{a}o Carlos,
S\~ao Paulo, Brazil\\
$^2$National Institute of Science and Technology for Complex systems, Brazil}

\begin{abstract}
An entropy-based image segmentation approach is introduced and applied
to color images obtained from \emph{Google Earth}. Segmentation refers
to the process of partitioning a digital image in order to locate
different objects and regions of interest. The application to
satellite images paves the way to automated monitoring of ecological
catastrophes, urban growth, agricultural activity, maritime pollution,
climate changing and general surveillance.  Regions representing
aquatic, rural and urban areas are identified and the accuracy of the
proposed segmentation methodology is evaluated. The comparison with
gray level images revealed that the color information is fundamental
to obtain an accurate segmentation.
\end{abstract}


\maketitle

\section{Introduction}

Medical, biological and astronomical experiments, as well as satellite
prospection, have generated terabytes of image data, making automatic
analysis a fundamental resource for knowledge discovery. Image
analysis is based on the extraction of meaningful information and can
involve many steps, such as pre-processing (e.g. noise removing),
segmentation and characterization of the identified
objects~\cite{da2001shape}.  Particularly, the identification of the
types of objects --- a task called \emph{segmentation} --- constitutes
an essential issue in pattern recognition~\cite{da2001shape} due to
its practical importance, such as in the treatment of images obtained
from satellite prospection. In fact, image segmentation can be
understood as the process of assigning a label to every pixel in an
image, such that pixels with the same label represent the same object,
or its parts

In the current work, we propose an entropy-based segmentation of
images. The methodology is evaluated with respect to satellite images
obtained from \emph{Google Earth}, in order to identify aquatic, urban
and rural regions. The importance of using
\emph{Google Earth} images can be observed in a growing number of
investigations, such as the analysis of magnetic alignment of cattle
and deer during grazing and resting~\cite{begall105mag} or mapping of
disaster zones for identifying priorities, planning logistics and
definition of access routes for relief
operations~\cite{nourbakhsh2006mdz}. In fact, satellite images are
critically important for the monitoring of ecological catastrophes,
urban growth, agricultural activity, maritime pollution, climate
changing as well as general surveillance. Moreover, the segmentation
of \emph{Google Earth} images is particularly important for automatic
mapping of urban and rural areas while monitoring dynamical human
activities, such as city growth that can affect regions of
environmental preservation. Another application involves monitoring of
rural activities, which can also lead to different textures, such as
those observed in cultivation of sugarcane or wheat. The
identification of aquatic areas allows the monitoring of pollution,
which can be potentially inferred from changes in the water texture,
as well as the formation of deserts or marshes, i.e.\ it provides an
indication about possible climate changes.  In addition, analysis of
satellite images can help in monitoring of deforestation and in
finding focuses of fires in forests.

Images are composed by a set of pixels whose values encode different
colors or gray levels. Image segmentation methods have been used to
find regions of interest (e.g. objects) in images. The importance of
image segmentation can be illustrated in diverse practical
applications, such as in medical imaging
(e.g. diagnosis~\cite{noble2006uis}), satellite
images~\cite{masson1993sem}, face recognition~\cite{zhao2003face},
traffic control system~\cite{cucchiara2000image} and machine
vision~\cite{haralick1992computer}. Different algorithms have been
proposed for image segmentation such as those founded on image
thresholding (e.g. by means of histograms of gray
levels~\cite{navon2005cis}); clustering methods (e.g neural
networks~\cite{kuntimad1999pis}); region growing methods
(e.g.~\cite{haralick1985ist}); graph partitioning methods
(e.g.~\cite{shi2000nca}); multi-scale segmentation
(e.g.~\cite{kuijper2003ssh}), and semi-automated segmentation
(e.g.~\cite{mortensen1998isi}). Methods related to physics concepts
have also been more and more applied for image segmentation, such as
those based on Markov random fields~\cite{celeux2003pum} and
entropy~\cite{portesdealbuquerque2004itu}.  The segmentation approach
proposed in the current work is based on the concept of entropy.

In next sections, the concepts of information entropy, dimensionality
reduction and supervised classification are presented. Afterwards, the
proposed image segmentation methodology is applied to images of the
\emph{Google Earth} and the classification results are evaluated. The
influence of the parameters involved in the segmentation is
discussed. Venues for future research and conclusions are identified.

\section{Methodology}

In information theory, the concept of entropy is used to quantify the
amount of information necessary to describe the macrostate state of a
system~\cite{cover1991eit}.  Therefore, the entropy is related to the
concept of complexity~\cite{bar2003dynamics}. Then, if a system
presents a high value of entropy, it means that much information is
neccessary to describe its states. Depending of the specific
application, the entropy can be defined in different ways. For
instance, while in quantum mechanics, the entropy is related to the
von Neumann entropy~\cite{vonneumann1932mgq}; in complexity theory, it
is associated to the Kolmogorov entropy~\cite{li1997ikc}. Here we take
the concept of entropy in the sense of information theory (Shannon
entropy), where entropy is used to quantify the minimum descriptive
complexity of a random variable~\cite{cover1991eit}. The Shannon
entropy of a discrete random distribution $p(x)$ is defined as
\begin{equation}
H(p) = -\sum_{x}p(x) \log p(x),
\end{equation}
where the logarithm is taken on the base 2.

In image analysis, $p(x)$ can refer to the distribution of gray levels
or to the intensity of different color components of an image. The
histograms $p(x)$ of a color image are obtained by counting the number
of pixels with a given color intensity (red (R), green (G) or blue
(B)), which can vary from 0 to 255. In this way, this procedure
generates a set of three different histograms $\{h_c(x)\}$, where $c =
\{R,G,B\}$. Due to its particular nature, as discussed above, the
entropy can provide a good level of information to describe a given
image. In this case, if all pixels in an image have the same gray
level or the same intensity of color components, this image will
present the minimal entropy value. On the other hand, when each pixel
of an image presents a specific gray level or a color intensity, it
this image will exhibit maximum entropy. Thus, since the pixel
intensities are related to texture, because different textures tend to
result in different distribution of gray level or color intensity, the
Shannon entropy can be used for texture
characterization~\cite{da2001shape}.  Our texture approach is based on
this assumption about texture analysis.  The application to satellite
images is justified because these images are formed by objects
presenting different textures.  In fact, different regions in these
images, such as aquatic and urban areas, tend to present specific
textures which are possibly characterized by different entropy values.
For instance, while urban areas tend to exhibit high color variations
(higher entropy), aquatic regions tend to be more homogeneous (lower
entropy).

Our proposed methodology for segmentation of satellite images is
performed as follows.  Images are divided into square windows with a
fixed size $L$, the entropy is calculated for each window, and then a
classification methodology is applied for the identification of the
category of the respective windows (e.g. aquatic, rural, urban,
etc.). The classification approach can be supervised or
non-supervised. Supervised classification needs a training set
composed by windows whose classes are previously known (prototypes),
such as rural and urban areas. Here, we focus on a segmentation
methodology based on supervised classification.  Initially, the
training is done by selecting samples (windows) of the three types of
regions (i.e.\ aquatic, rural and urban areas). Observe that each of
these sample windows should be selected in order to present pixels of
only one class. Next, the entropy is calculated for each color
component (R,G and B) of these windows. Therefore, these windows are
represented in a three-dimensional space defined by the entropy of the
colors components, i.e.\ each window is represented by a vector with
three elements. Then, due to the high correlation between the entropy
of color components, these windows are projected into a one
dimensional space by considering principal component
analysis~\cite{jolliffe2002principal}. Note that the projection into
one dimension by principal component analysis allows to optimally
remove the redundancy present in the data. Finally, the classification
of the training set is performed.

The classification is done by maximum likelihood decision theory,
which considers the density functions estimated for each
class~\cite{duda2001pattern}.  This estimation is obtained by the
Parzen windows approach~\cite{duda2001pattern}, which adds a
normalized Gaussian function at each observation point, so that the
interpolated densities correspond to the sum of these functions,
performed separately for each class (see
Figure~\ref{Fig:parzen}). These densities are used in the maximum
likelihood approach. If the probability density is known, it can be
showed that this classification approach is optimal in the sense of
minimizing misclassification~\cite{duda2001pattern}. The second step
in the supervised classification is performed by classifying unknown
windows. In this way, it is possible to evaluate the accuracy of the
classifier by comparing the resulting classification and the original
regions. In fact, the evaluation of the precision of the
classification approach is given by the confusion matrix $C$, whose
elements $c_{ij}$ provide the number of windows of class $j$ which
were classified as being of class $i$~\cite{da2001shape}. The
percentage of correct classification is obtained by the sum of the
confusion matrix diagonal divided by the total sum of the matrix.

\section{Results and discussion}

\begin{figure*}[!ht]
\begin{center}
 \subfigure[]{\includegraphics[width=0.3\linewidth]{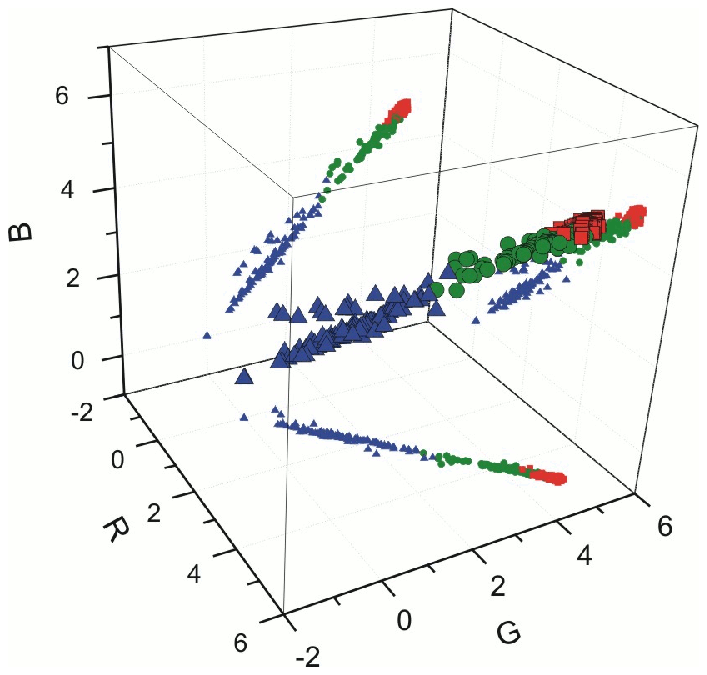}}\hspace{0.5cm}
 \subfigure[]{\includegraphics[width=0.3\linewidth]{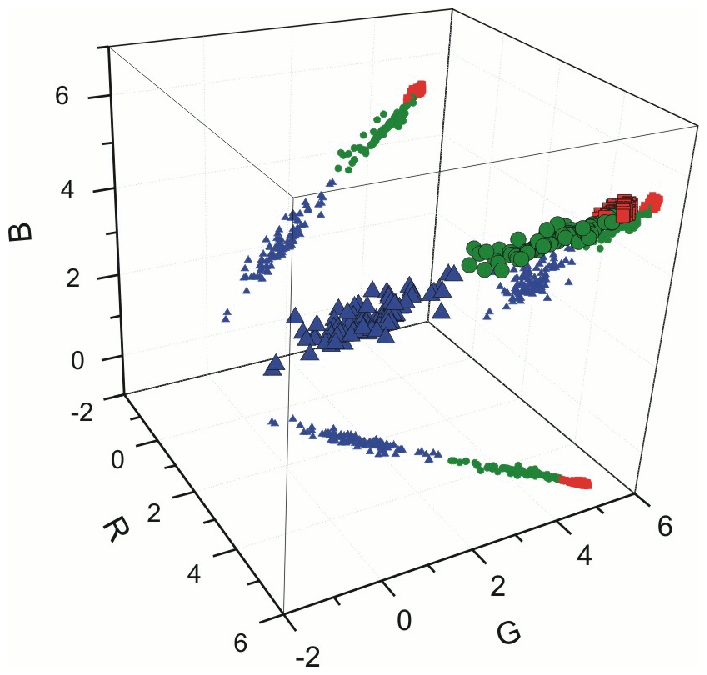}}\hspace{0.5cm}
 \subfigure[]{\includegraphics[width=0.3\linewidth]{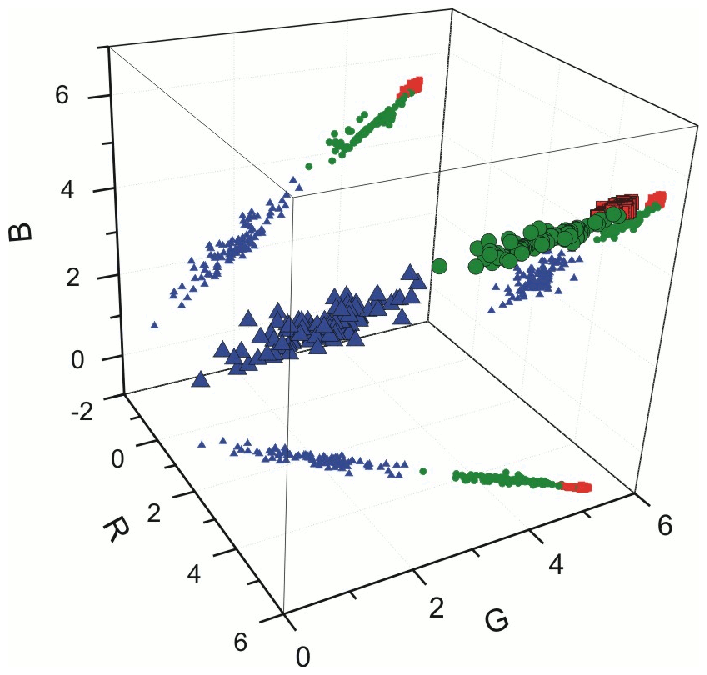}}
 \caption{The scatterplot of the entropies of the windows with sizes
 (a) $16\times 16$, (b) $30\times30$ and (c) $46\times46$. Each point
corresponds to a
 window, represented by the three coordinates associated to the
 entropies of each of the three color components (R, G and B).
Windows corresponding to
 water, rural and urban regions are represented by blue triangle,
green circles and red squares, respectively.
 These windows correspond to the training step of the supervised
classification.}~\label{Fig:windows}
 \end{center}
\end{figure*}

In order to segment images of the \emph{Google Earth}, we took into
account square windows of dimensions $16\times16$, $30\times30$ and
$46\times46$ pixels. We obtained $100$ windows of each class and
calculated the entropy distribution for each color component from the
respective histograms. Figure~\ref{Fig:windows} presents the windows
in the space defined by the entropy of the three color components.
Note that the urban and rural regions present a small intersecting
region, because urban areas can exhibit trees and parks, which present
textures similar to those present in rural areas.  Since these windows
are approximately organized as a straight line in the
three-dimensional scatterplot, which indicates a strong correlation
between the entropies of color components, we projected the entropies
into a one-dimensional space by applying principal component
analysis~\cite{jolliffe2002principal}. The variances of this type of
projection corroborate the one-dimensional organization of the points,
i.e.\ the first eigenvalue divided by all eigenvalues is equal to
$\lambda_1/\sum_{i=1}^3 \lambda_i = 0.99$ for all windows sizes.  In
other words, the projected data accounts for $99\%$ of the variance of
the original observations.  To obtain the density function, we
considered the Parzen windows approach, as described before.

\begin{figure*}[t]
\begin{center}
\subfigure[]{\includegraphics[width=0.3\linewidth]{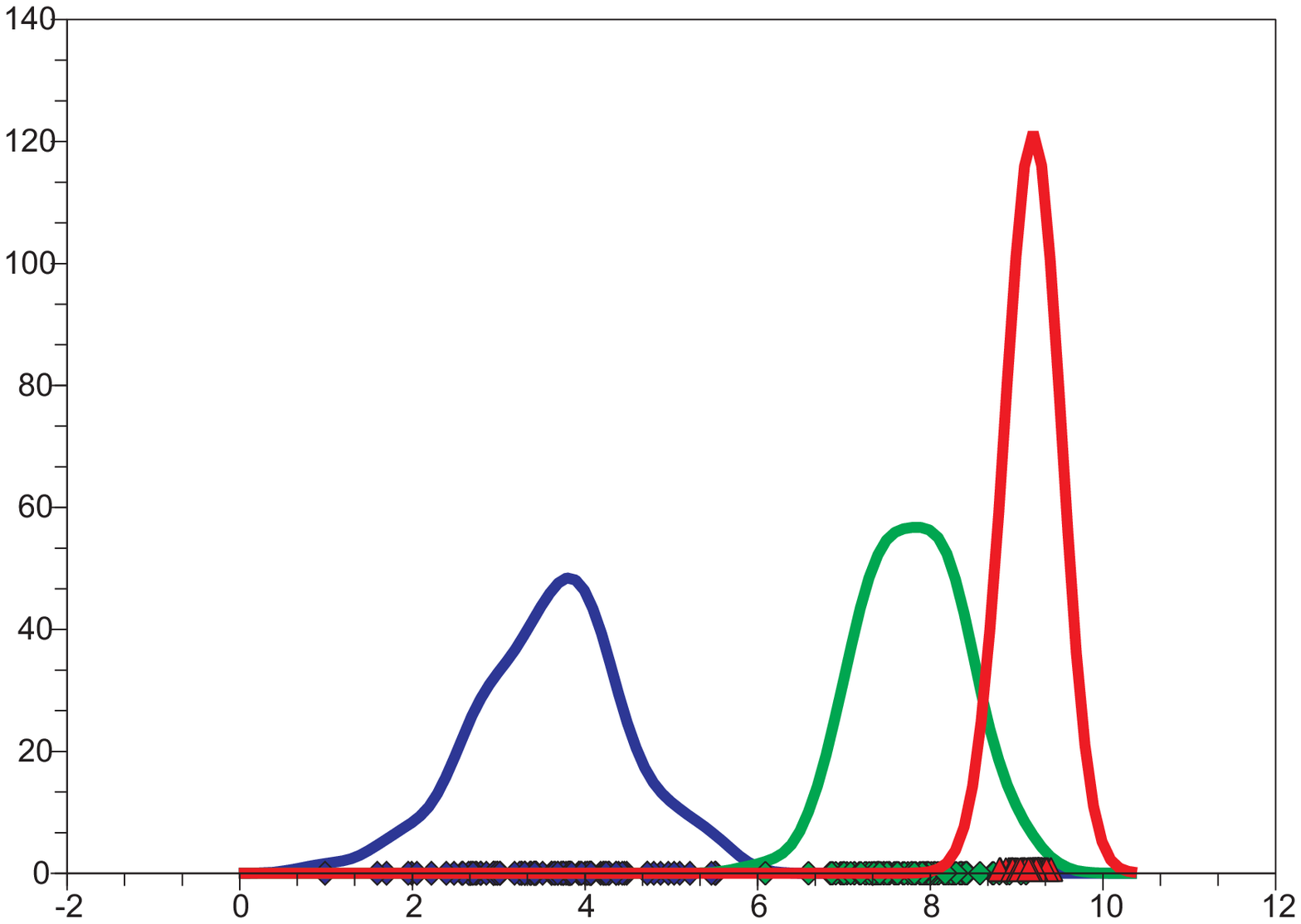}}\hspace{0.5cm}
\subfigure[]{\includegraphics[width=0.3\linewidth]{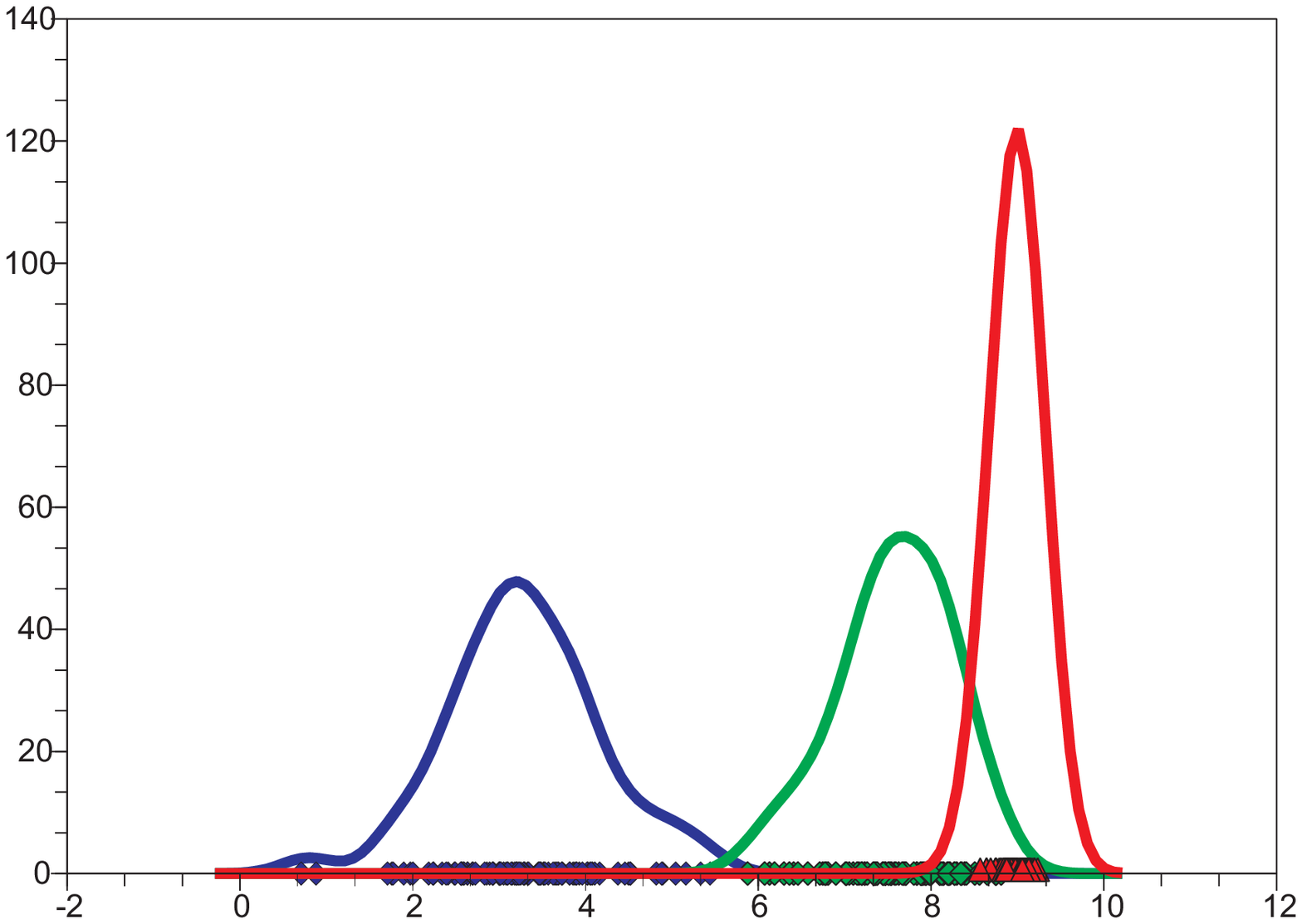}}\hspace{0.5cm}
\subfigure[]{\includegraphics[width=0.3\linewidth]{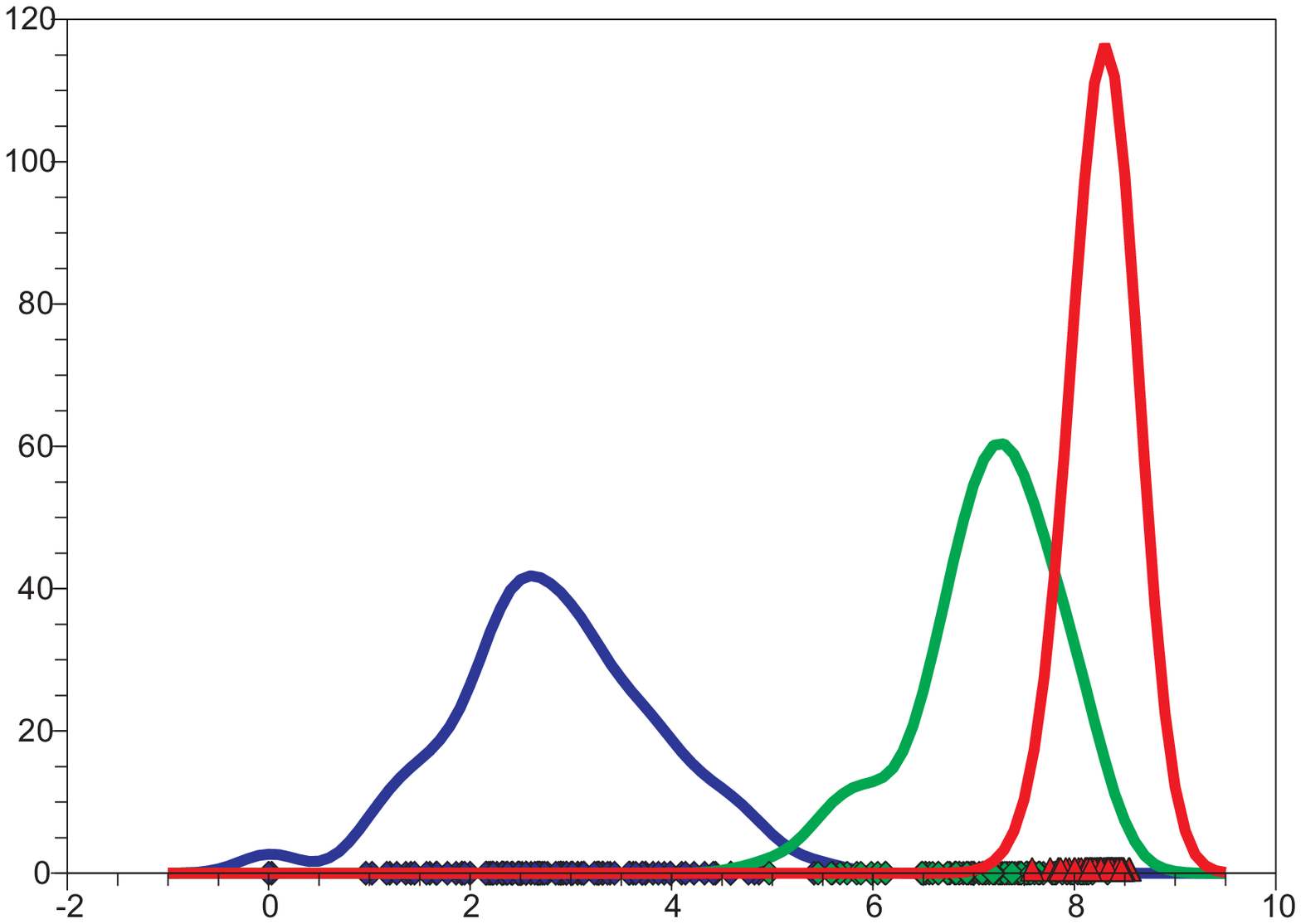}}
\end{center}
 \caption{Probability densities estimated for windows of size (a) 16, (b) 30
 and (b) 46. The projections were obtained from the scatterplots of
 Figure~\ref{Fig:windows}.  The windows correspond to
 water, rural and urban regions are represented by
 blue triangle, green circles and red squares, respectively.}
\label{Fig:parzen}
\end{figure*}

Figure~\ref{Fig:parzen} illustrates the obtained probability
densities. After estimation, we performed the classification by
maximum likelihood decision theory, which uses the Bayes rule,
associating each image window to the class that results in the largest
probability~\cite{duda2001pattern}.  Figure~\ref{Fig:parzen} shows
that the larger the windows sizes, the larger are the intersections
between the curves. In addition, urban and rural areas present the
largest intersecting region, because some urban areas are composed by
trees, woods and parks.

In order to evaluate the precision of our methodology, we segmented
$10$ images manually and compared these original segmentations with
those obtained from our classification methodology. The regions were
extracted from cities from different worldwide zones, such as Berlin,
Hong Kong, New York, Buenos Aires, Washington, Warsaw, Madrid, and
Baghdad. The images were obtained at the same altitude ($2,000$
meters), in order to incorporate the same level of details in each
sample. Tables~\ref{Tab:16}, \ref{Tab:30} and~\ref{Tab:46} present the
confusion matrices. Notice that these matrices were calculated by
taking into account each pixel on the image, and not each window,
because some windows are composed by more than one class of
pixels. The adoption of small windows, i.e.\ $16\times16$ and
$30\times30$, accounted to a more accurate classification than the
larger one ($46\times 46$). This happens because small windows tend to
include regions with more homogeneous classes, while more
heterogeneous regions tend to be included within larger
windows. Nevertheless, the precision obtained with smaller windows is
achieved at the expense of higher computational cost, due to the
larger number of required windows to be processed. Comparing the
percentage of correct classification given in each confusion matrix,
we conclude that the highest errors occurs for the aquatic and rural
regions with respect to windows of size $46 \times 46$, where $24\%$
of aquatic regions were classified as rural regions, and $24\%$ of
rural regions were classified as urban.  In the former case, the error
has been verified to be a consequence of texture similarities between
some rivers that present a high level of green algae and some types of
plantations, which are predominantly based on green vegetables. In the
latter case, urban and rural regions tend to share similar green areas.
The highest score ($90\%$) was obtained by aquatic regions taking into
account windows of size $16\times16$. The accuracy of our
classification methodology can be summarized in terms of the sum of
the confusion matrix diagonal divided by the total sum of the
matrix. We indicate this ratio by $\alpha_L$, where $L$ is the window
size. The obtained values are equal to $\alpha_{46} = 0.79\pm0.07$ for
windows of size $46 \times 46$, $\alpha_{30} = 0.85\pm0.03$ for
windows of size $30\times 30$ and $\alpha_{16} = 0.85\pm0.04$ for
windows of size $16\times 16$. Therefore, the smallest windows provide
the most precise segmentation.

\begin{table}[!ht]
\begin{center}
\begin{small}
\caption{Confusion matrix for $10$ segmented images taking into
account windows of size $16\times 16$.
The overall accuracy in this case is equal to $\alpha_{16}=0.85\pm0.04$.}
\begin{tabular}{l|c|c|c}
\hline Confusion &urban           &rural           &aquatic \\\hline
urban            &0.83$\pm0.07$   &0.17$\pm0.07$   &0.00$\pm0.00$ \\
rural            &0.17$\pm0.09$   &0.82$\pm0.09$   &0.01$\pm0.03$ \\
aquatic          &0.00$\pm0.00$   &0.10$\pm0.06$   &0.90$\pm0.06$ \\
\hline
\end{tabular}
\label{Tab:16}
\end{small}
\end{center}
\end{table}

\begin{table}[!ht]
\begin{center}
\begin{small}
\caption{Confusion matrix for $10$ segmented images taking into
account windows of size $30\times 30$.
The overall accuracy in this case is equal to $\alpha_{30}=0.85\pm0.03$.}
\begin{tabular}{l|c|c|c}
\hline Confusion &urban           &rural           &aquatic \\\hline
urban            &0.87$\pm0.05$   &0.13$\pm0.05$   &0.00$\pm0.00$ \\
rural            &0.13$\pm0.07$   &0.86$\pm0.06$   &0.01$\pm0.02$ \\
aquatic          &0.01$\pm0.01$   &0.17$\pm0.10$   &0.82$\pm0.11$ \\
\hline
\end{tabular}
\label{Tab:30}
\end{small}
\end{center}
\end{table}

\begin{table}[!ht]
\begin{center}
\begin{small}
\caption{Confusion matrix for 10 segmented images taking into account
windows of size $46\times 46$.
The overall accuracy in this case is equal to $\alpha_{46}=0.79\pm0.07$}
\begin{tabular}{l|c|c|c}
\hline Confusion &urban       &rural        &aquatic \\\hline
urban            &0.88$\pm0.06$    &0.11$\pm0.06$   &0.01$\pm0.01$ \\
rural            &0.24$\pm0.12$    &0.75$\pm0.11$   &0.01$\pm0.02$ \\
aquatic          &0.02$\pm0.02$    &0.24$\pm0.16$   &0.74$\pm0.17$ \\
\hline
\end{tabular}
\label{Tab:46}
\end{small}
\end{center}
\end{table}

\begin{figure}[t]
\begin{center}
\subfigure[]{\includegraphics[width=0.9\linewidth]{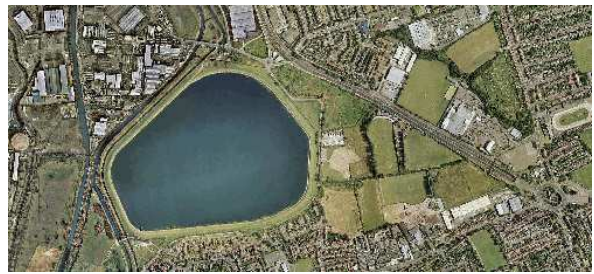}}\\
\subfigure[]{\includegraphics[width=0.9\linewidth]{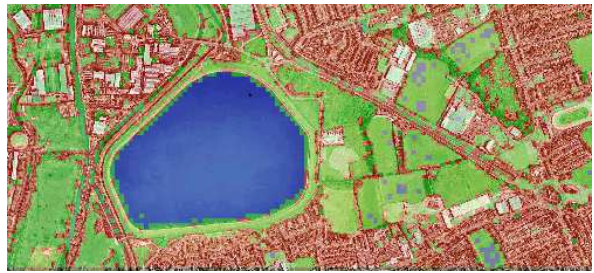}}\\
\subfigure[]{\includegraphics[width=0.9\linewidth]{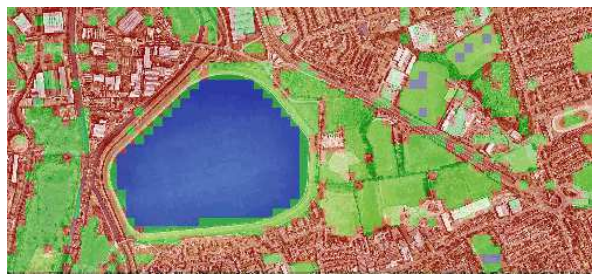}}\\
\subfigure[]{\includegraphics[width=0.9\linewidth]{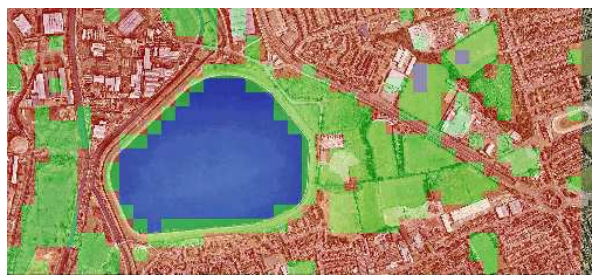}}
\end{center}
 \caption{A region of London and its respective segmentation by
taking into account windows of size (a) 16, (b) 30 and (c) 46.
 } \label{Fig:london}
\end{figure}

An additional analysis of our classification methodology was performed
with respect to the segmentations of a region of London (obtained at
$2,000$ meters of altitude), as presented in Figures~\ref{Fig:london}
for windows of dimensions $16\times 16$, $30\times 30$ and $46\times
46$. The smallest windows ($16\times16$) provide the most accurate
segmentation, mainly with respect to the boundaries of the rural,
aquatic and urban regions. Nevertheless, at the same time, due to the
small size of the windows, some parts of urban areas are classified as
rural as a consequence of the presence of trees, woods and parks. In
fact, due to the level of details of the image, some windows
corresponding to urban areas can be completely formed by trees --
windows composed by green areas typically correspond to rural
regions. As we increase the size of the windows, the observed
misclassification is reduced, but the boundaries of each region tend
to become less defined. This effect can be observed along the boundary
of the aquatic area. Indeed, the effect of the green regions in urban
area segmentation can be verified by the comparison of the confusion
matrices obtained for windows of size $16\times16$ and $30\times30$,
Tables~\ref{Tab:16} and~\ref{Tab:30}. These tables show that the
former case results in a larger error in classification of urban
regions, mainly due to the classification of urban trees as rural
areas. These misclassifications implied in similar scores for windows
of dimensions $16\times16$ and $30\times30$. Indeed, the more accurate
segmentation of the boundary of the regions are compensated by the
wrong classification of urban green areas. Despite the wrong
segmentation of these areas, we can observe that more accurate
classifications can be obtained for smaller windows. Larger windows
tend to provide worse classification because many of these windows in
the segmented image can be compose by more than one class of
regions. In fact, most of the misclassifications occur with respect
to these windows.

In order to compare our obtained results with a more traditional
approach, we took into account gray level versions of the considered
images.  We adopted the same methodology used for color images to
obtain the segmentation, but each image was now represented by a
vector with only one element (the entropy of gray level
histograms). Note that for color images, three color components were
used and the images were represented by a vector composed by three
elements.  Tables~\ref{Tab:16_gray}, \ref{Tab:30_gray} and
\ref{Tab:46_gray} show the obtained confusion matrices for windows of
dimensions $16\times 16$, $30\times 30$ and $46\times 46$,
respectively. In these cases, the sum of the confusion matrix diagonal
divided by the total sum of the matrix are equal to $\alpha_{16} =
0.74\pm0.10$, $\alpha_{30} = 0.75\pm10$ and $\alpha_{46} =
0.73\pm0.10$.  It is nteresting to observe that the different windows
sizes resulted in similar classification performances.  Comparing with
the results obtained for color images, the gray level resulted in
worse classification. Therefore, the spectral color information is
critically important for achieving accurate segmentation.

\begin{table}[!ht]
\begin{center}
\begin{small}
\caption{Confusion matrix for $10$ segmented gray images taking into
account windows of size $16\times 16$. The overall accuracy in this
case is equal to $\alpha_{16}=0.74\pm0.10$}
\begin{tabular}{l|c|c|c}
\hline Confusion &urban           &rural           &aquatic \\\hline
urban            &0.74$\pm0.20$   &0.26$\pm0.20$   &0.00$\pm0.00$ \\
rural            &0.11$\pm0.09$   &0.89$\pm0.09$   &0.00$\pm0.01$ \\
aquatic          &0.02$\pm0.05$   &0.38$\pm0.28$   &0.60$\pm0.27$ \\
\hline
\end{tabular}
\label{Tab:16_gray}
\end{small}
\end{center}
\end{table}

\begin{table}[!ht]
\begin{center}
\begin{small}
\caption{Confusion matrix for $10$ segmented gray images taking into
account windows of size $30\times 30$. The overall accuracy in this
case is equal to $\alpha_{30}=0.75\pm0.10$}
\begin{tabular}{l|c|c|c}
\hline Confusion &urban           &rural           &aquatic \\\hline
urban            &0.74$\pm0.22$   &0.26$\pm0.22$   &0.00$\pm0.00$ \\
rural            &0.09$\pm0.09$   &0.91$\pm0.08$   &0.01$\pm0.01$ \\
aquatic          &0.01$\pm0.02$   &0.39$\pm0.26$   &0.59$\pm0.26$ \\
\hline
\end{tabular}
\label{Tab:30_gray}
\end{small}
\end{center}
\end{table}

\begin{table}[!ht]
\begin{center}
\begin{small}
\caption{Confusion matrix for 10 segmented gray images taking into
account windows of size $46\times 46$. The overall accuracy in this
case is equal to $\alpha_{46}=0.73\pm0.10$}
\begin{tabular}{l|c|c|c}
\hline Confusion &urban       &rural        &aquatic \\\hline
urban            &0.75$\pm0.22$   &0.24$\pm0.22$   &0.01$\pm0.01$ \\
rural            &0.10$\pm0.10$   &0.89$\pm0.10$   &0.01$\pm0.01$ \\
aquatic          &0.02$\pm0.04$   &0.39$\pm0.26$   &0.59$\pm0.25$ \\
\hline
\end{tabular}
\label{Tab:46_gray}
\end{small}
\end{center}
\end{table}

\section{Conclusion}

Despite its simplicity, the described methodology revealed to be
particularly accurate and effective for the classification of
geographical regions.  Indeed, we have shown that the entropy of the
color distribution in images of geographical regions conveys enough
information about the respective type of terrain so as to ensure a
particularly high number of correct classifications, making of the
proposed methodology an operational approach to be used in several
related problems.  Although the best classification rate obtained was
equal to $0.90$, more accurate classification could be obtained by
taking into account windows of smaller sizes than those we used
here. Other statistical measurements, such as statistical moments, can
also be used to complement the characterization of the texture of
geographical regions. The extension of the current methodology to
other types of regions, such as different types of forest or
agricultural activities, is straightforward. In addition, the
classification methodology can be improved by considering smaller
windows combined with image pre-processing techniques, such as color
equalization or noise removal. Other types of classifiers, such as
support vector machine or neural networks~\cite{bishop2006pattern} can
also be used.

\section{Acknowledgments}

Luciano da F. Costa is grateful to FAPESP (proc. 05/00587-5), CNPq
(proc. 301303/06-1) for financial support.  Francisco A. Rodrigues
acknowledges FAPESP sponsorship (proc. 07/50633-9). Odemir M. Bruno acknowledges sup
port from CNPq (306628/2007-4 and 484474/2007-3).

\bibliographystyle{unsrt}
\bibliography{paper}

\end{document}